\documentclass[prl,twocolumn,showpacs,amsmath,amssymb,floatfix]{revtex4}

\usepackage{graphicx}
\usepackage{dcolumn}

\newcommand{\e}{\varepsilon}
\def \be{\begin{equation}}
\def \ee{\end{equation}}
\def \bea{\begin{eqnarray}}
\def \eea{\end{eqnarray}}
\def \be{\begin{equation}}
\def \ee{\end{equation}}
\newcommand{\te}[1]{\mbox{\boldmath $ #1 $}}
\newcommand{\tes}[1]{\mbox{\boldmath ${\it #1 }$}}
\newcommand{\rd}{\mathrm{d}}            

\begin{document}

\title{How efficient is towing a cargo by a micro-swimmer?}
\author{O. Raz$^1$} \author{A. M. Leshansky$^2$}\email{lisha@tx.technion.ac.il}

\affiliation{$^1$Department of Physics and $^2$Department of Chemical Engineering \\
Technion--IIT, Haifa, 32000, Israel \\}

\date{\today}

\begin{abstract}
We study the properties of arbitrary micro-swimmers towing a passive load through a viscous liquid. The simple close-form expression for the dragging efficiency of a general micro-swimmer dragging a distant load is found, and the leading order approximation for finite mutual separation is derived. We show that, while swimmer can be arbitrarily efficient, dragging efficiency is always bounded from above. It is also demonstrated, that opposite to Purcell's assumption, the hydrodynamic coupling can ''help" the swimmer to drag the load. We support our conclusions by rigorous numerical calculations for the ``necklace-shaped" swimmer, towing a spherical cargo positioned at a finite distance.
\end{abstract}

\maketitle

In the recent years there has been an increasing interest in propulsion at low Reynolds numbers, both theoretically \cite{Stone-Samuel,Avron-Gat-Kenneth,Wilczek-Shapere,JFM-Stone,N-Linked-Rods,PushMePullYou,Hosoi,Golestanian,treadmilling,wada,Pumping,TankTreading} and experimentally \cite{Nature,Pnas-swimming-efficiency,MagneticallyDerivenSwimmer}. These and other works have improved our understanding of the basic properties of locomotion on small scales. However, it is not sufficient to understand the mechanisms and properties of free microswimmers alone -- it is necessary to estimate the performance of these swimmers as \emph{propellers} that tow a useful cargo, e.g. therapeutic load or miniature camera. This question, which attracted only limited attention, had already shown to have some non-trivial answers: E.~M. Purcell had studied \cite{Purcell_helix} the particular case of a rotating helix pushing a spherical particle under the assumption of negligible hydrodynamic interactions. He showed that due to the structure of grand-resistance matrix, which connects the force and torque on a body to its translational  and angular velocities, the optimal rotating--propeller should have the same size as the load. 

In this paper we address arbitrary shaped swimmers and cargoes, and investigate the effect of their mutual hydrodynamic interaction on performance of the swimmer as load propeller. We find that, while propellers that can enclose a load within may theoretically have arbitrarily high efficiency (consider, for instance, the ``treadmiller" \cite{treadmilling}), the dragging efficiency of a swimmer towing a remote load is always bounded from above, and there is an optimal propeller/load size ratio. We also find, in contrast to Purcell's assumption \cite{Purcell_helix}, that in many cases the hydrodynamic coupling between the load and the propeller enhances the dragging efficiency. We support our theory by numerical calculations for a necklace-shaped swimmer \cite{TankTreading} towing a spherical cargo.

We start our analysis by considering a micro-swimmer (i.e. a propeller) dragging a distant load. In this case, we can neglect the mutual hydrodynamic interaction and calculate the dragging efficiency as 
\begin{equation}\label{eq:DragEff}
\varepsilon_d=\frac{K_l V_{d}^{2}}{P_d}
\end{equation}
where $K_l$ is the resistance coefficient of the load \footnote{We treat $K_l$ as a scalar, since the resistance matrix is symmetric \cite{Happel-Brenner}, and it is optimal to drag the load when the velocity is aligned in the direction of the minimal eigenvalue}, $V_d$ is the dragging velocity and $P_d$ is the power expended by the swimmer to drag the load with velocity $V_d$. We also define the propeller's efficiency in the same fashion, $\varepsilon_s=\frac{K_s V_{s}^{2}}{P_s}\:$, where $K_s$ is the swimmer's resistance coefficient, $V_s$ is the speed of the unloaded propeller (at the point where the load is anchored) and $P_s$ is the power expended in swimming without load.

Note that in a general case of swimmer propelled by a sequence of geometrically non-reciprocal periodic strokes (e.g. three-link Purcell's swimmer \cite{Purcell, JFM-Stone, Hosoi}, pushmepullyou \cite{PushMePullYou}, three-sphere swimmer \cite{Golestanian}, shape deformations \cite{Stone-Samuel,Wilczek-Shapere2} and others), the swimming efficiency is conventionally defined using stroke-averaged quantities \cite{Lighthill}. However, since $\mbox{max}\left\{\frac{K_s V_{s}^{2}}{P_s}\right\}>\frac{K_s\langle V_s\rangle^2}{\langle P_s \rangle}$ (where $\langle\,\rangle$ stands for average over a stroke period; the maximum is taken over the stroke period), the maximum of (\ref{eq:DragEff}) over a stroke period is an upper bound for the conventional efficiency. In the case of swimmer propelled without the shape change (e.g. rotating flagella \cite{Purcell_helix}, treadmiller \cite{treadmilling}, twirling torus \cite{TankTreading}, and others), the two definitions coincide. They are also practically equivalent for swimmers performing small-amplitude strokes, with $K_s \approx \mbox{Const}$ \cite{Stone-Samuel,Wilczek-Shapere2}. Also, note that (\ref{eq:DragEff}) is not just the swimming efficiency re-written for ``swimmer+load" as a new swimmer, since we aim to compare the power expended in dragging the load alone by swimmer and by an external force.

\begin{figure}
\begin{center}
  \includegraphics[scale=0.8]{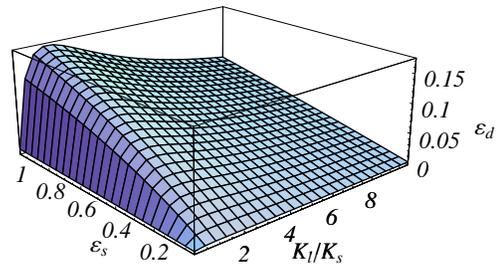}\\
  \caption{(Color online) The dragging efficiency, $\varepsilon_d$, as function of the propeller's efficiency $\varepsilon_s$ and the size ratio $K_l/K_s$}\label{fig:EffLargDist}
\end{center}
\end{figure}

We will now calculate the dragging efficiency for swimmer characterized by a resistance coefficient $K_s$ and swimming efficiency $\varepsilon_s$, dragging a load characterized by a resistance coefficient $K_l$, which we will assume are both not rotating (it is known \cite{Stone-Samuel} that a rotating swimmer is less efficient than a non-rotating one). By Lorentz reciprocity \cite{Happel-Brenner}, if $(v_j,\sigma_{jk})$ and $(v_j', \sigma_{jk}')$ are the velocity and stress fields for two solutions of the Stokes equations $\partial_j\sigma_{ij}=0$ in fluid domain $\Sigma$ then
\begin{equation}\label{eq:Lorenz reciprocity}
\int_{\partial \Sigma} v'_i\, \sigma_{ij}\, \rd S_j=\int_{\partial \Sigma} v_i\,\sigma_{ij}'\, \rd S_j
\end{equation}
Using (\ref{eq:Lorenz reciprocity}) with $(v_i,\sigma_{ij})$ being the velocity and the stress fields for a swimmer dragging a load, and $(v'_i,\sigma'_{ij})$ being the velocity and stress fields for the ``unloaded" swimmer and the load co-dragged by the external force with the swimmer's velocity,  we readily obtain
\begin{equation}\label{eq:PowerDrag}
P_d=P_{s+l}-(V_s-V_d)\cdot F_{s+l}\:.
\end{equation}
Here $P_d$ is the power expended by the swimmer to drag the load, $P_{s+l}$ is the power dissipated by viscosity in the case of ``unloaded" swimmer and the load co-dragged by the external force, $V_s$ is the velocity of the free swimmer, $V_d$ is the dragging velocity and $F_{s+l}$ is the force required to tow the load with velocity $V_s$. $V_d$ can be found by equating the sum of the viscous drag forces on the swimmer and the load to zero. Exploiting the linearity of Stokes equation and neglecting hydrodynamic interaction, we get:
\begin{equation}\label{eq:DraggVel}
V_d=\frac{V_s K_s}{K_s+K_l}\:.
\end{equation}
As expected, $V_d$ goes to zero for infinity large load and to the swimmer velocity for a vanishingly small load.
Neglecting hydrodynamic interaction, we can use $F_{s+l}=-V_s\:K_l$ and $P_{s+l}=P_s+V_s^2\: K_l$, that together with (\ref{eq:DraggVel}) and dragging power reads
\begin{equation}\label{eq:DragPowerDist}
P_d=P_s+V_s^2\frac{K_l\: K_s}{K_l+K_s}\:.
\end{equation}
For small loads, $K_l\ll K_s$, Eq.(\ref{eq:DragPowerDist}) gives the power of the free swimmer plus the power of dragging the load, and for large loads, $K_l\gg K_s$, this gives the power of an anchored swimmer (i.e. a ``pump" \cite{Pumping}).
Substituting (\ref{eq:PowerDrag}), the swimmer efficiency and (\ref{eq:DraggVel}) into (\ref{eq:DragEff}) gives
\begin{equation}\label{eq:DragEffLargDist}
\varepsilon_d=\frac{r}{(r+1)\left(\frac{r+1}{\varepsilon_s}+r\right)}\:,
\end{equation}
where $r=K_l/K_s$. The dependence of the dragging efficiency $\varepsilon_d$ on $\varepsilon_s$ and $r$ is plotted in Fig.~(\ref{fig:EffLargDist}). Eq.~(\ref{eq:DragEffLargDist}) shows that unlike the swimming efficiency, which, for some swimmers, can be arbitrarily high \cite{PushMePullYou,treadmilling}, the dragging efficiency is bounded by $\varepsilon_d\leq \frac{1}{r+1}<1$ even for $\varepsilon_s=\infty$. This means that enclosing a cargo \emph{within} the swimmer can be much more efficient than towing a remote one, and that there is an optimal swimmer size for \emph{any} swimming technique (including swimming techniques in which $r$ is varying periodically). As one might expect, $\e_d$ is a growing function of $\e_s$. However, while for inefficient propeller (like a rotating helix) the optimal size is about the same as the load size, the efficient swimmer with $\varepsilon_s \gg 1$ (e.g. ``pushmepullyou" \cite{PushMePullYou}) will be efficient as propeller only if it is much larger then the load. Thus, the naive intuition saying that the swimmer's efficiency alone controls the dragging efficiency, is not always right: in some cases less efficient but larger propeller is advantageous.

Now let us estimate the effect of hydrodynamic interaction between the propeller and the passive cargo separated by distance $d$.  For finite separation distances it is no longer valid to assume that $F_{s+l}=-V_s\:K_l$. However, the force must still be linear in the dragging velocity and we can write $F_d=\lambda_l K_l\:V_d$. In the same way, the force on the swimmer must be proportional to the changes in the velocity, so we will denote it by $F_s=\lambda_s (V_s-V_d)K_s$. Since the forces must still sum up to zero, the dragging velocity is
\begin{equation}\label{eq:DragVelFixed}
V_d=\frac{V_sK_s}{K_s+\frac{\lambda_l}{\lambda_s}\:K_l}
\end{equation}
Comparing the velocity in (\ref{eq:DragVelFixed}) to that for infinite separation (\ref{eq:DraggVel}), it can be readily seen that the deviation between the two depends on the ratio $\frac{\lambda_l}{\lambda_s}$: if $\frac{\lambda_l}{\lambda_s}>1$ the velocity will be lower then that in (\ref{eq:DraggVel}), and if $\frac{\lambda_l}{\lambda_s}<1$ the velocity will be higher than the infinite distance case. Since generally $\lambda_s,\:\lambda_l<1$ \cite{Happel-Brenner}, and for asymmetric configurations the resistance coefficient of the larger object will be almost constant, we can conclude that \emph{a large swimmer will drag faster when positioned close to the load, while a small swimmer will drag faster when located far from the load}.

Assuming that the separation distance is large enough, so $d>\max\{R_l,R_s\}$, where $R_l$ and $R_s$ are the hydrodynamic radii of the load and the propeller, respectively, we can now estimate the power needed for the swimmer to drag the cargo: it is known \cite{Pumping} that for any swimmer $P_s=P_p-P_g$, where $P_p$ is the power needed by the pump, i.e. the anchored swimmer, $P_s$ is the power needed by the swimmer when it is swimming freely and $P_g$ is the power needed to drag a ``frozen" (immobile) swimmer with the swimming velocity. If we use this relation by treating the swimmer plus the load as a modified swimmer, we can estimate the power needed to drag the load. In this case, $P_g$ is just the power needed to drag both the load and the (frozen) swimmer with velocity $V_d$ provided by (\ref{eq:DragVelFixed}), which is $P_g=P_{g(s)}\frac{\lambda_s}{\frac{K_l}{K_s}\frac{\lambda_l}{\lambda_s}+1}$ where $P_{g(s)}=V_s^2 K_s$ is the power needed to drag the immobile swimmer. As an approximation, we will assume that the power expanded by the pump does not depend on the proximity of the load, since $P_p=P_{p(s)}+\mathcal{O}[(\frac{R}{d})^2]$ ($P_{p(s)}$ is the power expanded by the pump when the load is absent). Together, this gives
\begin{equation}\label{eq:PowerEst}
P_d=P_s+V_s^2K_s\left(1-\frac{\lambda_s}{\frac{K_l}{K_s}\frac{\lambda_l}{\lambda_s}+1}\right)
\end{equation}
where $P_s$ is the power expended by the swimmer when the load is absent. Obviously, $P_d \geq P_s$ and the equality holds only when $K_l=0$.

Substitution of (\ref{eq:PowerEst}) and (\ref{eq:DragVelFixed}) in (\ref{eq:DragEff}) yields
\begin{equation}\label{eq:EffEstFixed}
\varepsilon_d=\frac{r}{\left[\left(\frac{\lambda_l\: r}{\lambda_s}+1 \right)\left(1+\frac{1}{\varepsilon_s} \right)-\lambda_s \right]\left(\frac{\lambda_l\:r}{\lambda_s}+1\right)}\:,
\end{equation}
where $r=K_l/K_s$. For $\lambda_s=\lambda_l=1$ (\ref{eq:EffEstFixed}) reduces to (\ref{eq:DragEffLargDist}), as anticipated. Comparing the efficiency in (\ref{eq:EffEstFixed}) to that in (\ref{eq:DragEffLargDist}) one can conclude
that in cases where $\frac{\lambda_l}{\lambda_s}>1$ (i.e. the swimmer is smaller than the load), the efficiency is lower when the hydrodynamic coupling is not negligible, and it would better be separated from the load. If the swimmer is much bigger than the load, which implies $\lambda_s \simeq 1$ and $\frac{\lambda_l}{\lambda_s}<1$, the efficiency is higher than in the case with no coupling! Thus propeller bigger than the load should be positioned closer to the load, opposite to Purcell's assumption \cite{Purcell_helix}. Eq.~(\ref{eq:EffEstFixed}) also tells that the efficiency is bounded by $\frac{\lambda_s}{\lambda_l}$, which for large propeller towing a small load can theoretically be grater than 1. However, we could not find such an example.

We can now estimate $\lambda_s$ and $\lambda_l$ as functions of $d$, using the Oseen tensor \cite{Happel-Brenner}. As the first order approximation, we will assume both the swimmer and the load can be modeled as spheres with hydrodynamic radii $R_s=\frac{K_s}{6\pi\mu}$ and $R_l=\frac{K_l}{6\pi\mu}$, respectively \footnote{If either load or propeller deviates considerably from the spherical shape, it can be taken care of in the far-field resistance tensor. Here, for simplicity, we only refer to the far-field interaction between two spheres}. In this case, it can be readily shown that for $d\gg \max\{R_l,R_s\}$ \cite{Happel-Brenner}: $\lambda_s=\frac{2(2d^2-3 d R_l)}{4 d^2-9 R_l R_s}$, $\lambda_l=\frac{2(2d^2-3d R_s)}{4d^2-9 R_l R_s}$.
Substituting these expressions into (\ref{eq:DragVelFixed}) gives the expression for the dragging speed,
\be\label{eq:VelWithasak}
V_d=V_s\:\frac{2 \delta - 3 r}{2 (\delta - 3 r + \delta r)}\:,
\ee
where $\delta=d/R_s$ is the scaled separation distance. Similarly, substitution of $\lambda_l$ and $\lambda_s$ into (\ref{eq:EffEstFixed}) yields
\be\label{eq:EffWithasak}
\varepsilon_d=\frac{r}{\left[\left(\frac{2\delta-3}{2\delta-3r} r+1\right)\left(1+\frac{1}{\varepsilon_s}\right)
-\frac{2(2\delta^2-3r\delta)}{4\delta^2-9r}\right]\left(\frac{2\delta-3}{2\delta-3r} r+1\right)}.
\ee
Expanding (\ref{eq:EffWithasak}) for small $\frac{1}{\delta}$ gives
$\varepsilon_d \approx \varepsilon_{d(\infty)}+\frac{3\varepsilon_{d(\infty)}^2}{\varepsilon_s \delta} \left[1-(1+\varepsilon_s)r^2\right]+\ldots\:$,
where $\varepsilon_{d(\infty)}$ corresponds to the no-hydrodynamic-interaction approximation for the dragging efficiency (\ref{eq:DragEffLargDist}).
The $1/\delta$-term in the above expansion shows that for $r>\frac{1}{\sqrt{1+\varepsilon_s}}$ the dragging is retarded in comparison to the infinite separation result (\ref{eq:DragEffLargDist}), i.e. $\varepsilon_d<\varepsilon_{d(\infty)}$, while for $r<\frac{1}{\sqrt{1+\varepsilon_s}}$, the dragging is enhanced due to the hydrodynamic coupling, as $\varepsilon_d>\varepsilon_{d(\infty)}$. Interestingly, $r=\frac{1}{\sqrt{1+\varepsilon_s}}$ corresponds to the maximum of $\varepsilon_{d(\infty)}$. However, it is not the optimum of $\varepsilon_d$, which shifts to higher values at smaller $r$'s. 

\begin{figure}
\begin{center}
  \includegraphics[scale=0.3]{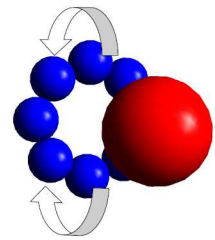}\\
  \caption{(Color online) Schematic of the necklace-like propeller towing a spherical load. The arrows show the direction of the rotation of spheres in the propeller; the propeller is pushing the load in front of it.}\label{fig:schematic}
\end{center}
\end{figure}

We shall now test the proposed theory for the load dragged by a rotary propeller. Imagine the necklace-like ring (see Fig.\ref{fig:schematic}) of $N_p=8$ nearly touching rigid spheres (separated by the distance of $0.05 a$) of radius $a$. The necklace lies in the $xy$ plane and in a cylindrical polar coordinate system $(z,r,\varphi)$, each sphere rotates at the constant angular velocity $\tes{\Omega}=\omega \te{e}_\varphi$, which, in the absence of external forces, causes the necklace to swim along the normal to the plane of the necklace in the positive $z$ direction \cite{TankTreading}. Performance of this swimmer as cargo propeller is tested for a spherical particle positioned at arbitrary distance along $z$-axis \footnote{For computational simplicity, the potential hydrodynamic disturbance caused by the links required to connect the propeller and the load is neglected.}. The distance that separates the plane of the propeller ($z=0$) and the load's surface is denoted  by $d_*$.

\begin{figure}[tb]
\begin{center}
 \includegraphics[scale=0.8]{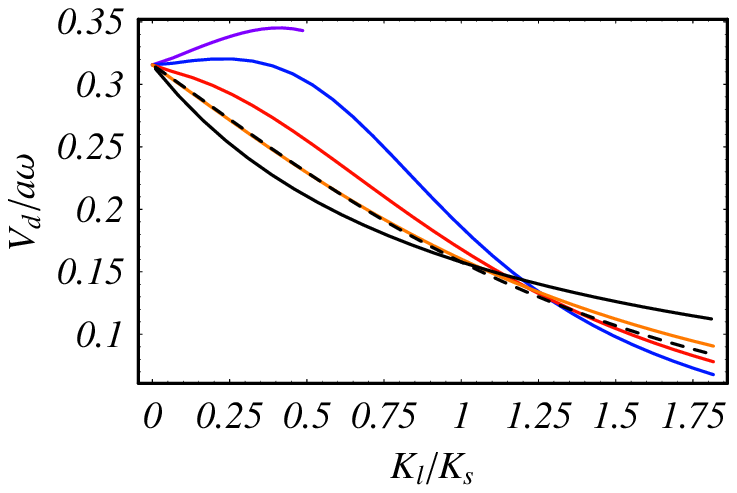}\\
 \includegraphics[scale=0.8]{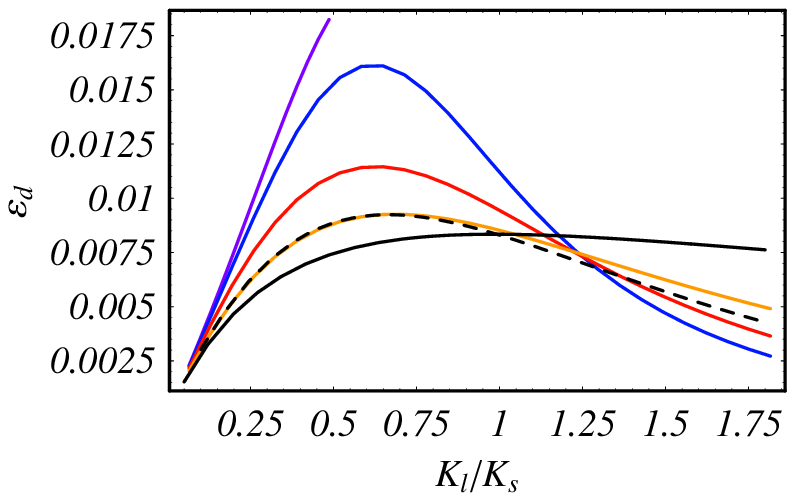}
  \caption{(Color online) Numerical results for the ``necklace-shaped" propeller made of 8 co-rotating spheres of radius $a$ ($V_s/a\omega\simeq0.316$, $\varepsilon_s=0.0339$, $R_s=3.083a$), towing a spherical load of variable size located at: $d_*=0$ (magenta), $d_*=a$ (blue), $d_*=4a$ (red) and $d_*=10a$ (yellow); the solid line corresponds to the infinite-separation result (\ref{eq:DraggVel}); the dashed lines are the far-field approximations for $d_*=10a$. (\emph{a}) the scaled dragging speed, $V_d/a\omega$; (\emph{b}) the dragging efficiency, $\varepsilon_d$.}\label{fig:Np8VelEff}
\end{center}
\end{figure}

We use the Multipole Expansion method \cite{TankTreading} and construct the rigorous solution of the Stokes equations as superposition of Lamb's spherical harmonic expansions \cite{Happel-Brenner}. The no-slip conditions at the surface of all spheres are enforced via the direct transformation between solid spherical harmonics centered at origins of different spheres. The accuracy of calculations is controlled by the number of spherical harmonics, $L$,  retained in the series. The truncation level of $L \le 7$ was found to be sufficient for all configuration to achieve an accuracy of less than 1\%. The dragging efficiency (\ref{eq:DragEff}) for this particular swimmer reads
\be\label{eq:DragEff1}
\e_d=\frac{K_l V_{d}^{2}}{N_p T\omega}\:,
\ee
where $\te{T}=\int_{S_i} \te{r}_i\times (\te{\sigma\cdot}\te{n})\:\rd S$ is a hydrodynamic torque exerted on each ($i$th) sphere of the propeller towing the load with $\te{r}_i$ being the radius vector with origin at the $i$th particle center. The values of $K_l$, $T$ and $V_d$ are determined numerically and the resulting scaled dragging speed $V_d/a\omega$, and efficiency, $\varepsilon_d$, are plotted vs. the size ratio $r$ in Figs. \ref{fig:Np8VelEff}(\emph{a}) and (\emph{b}), respectively. The agreement with the asymptotic results (\ref{eq:VelWithasak},\ref{eq:EffWithasak}) (via $\delta=d_*/R_s+r$) is excellent for small loads ($r<1$) even at moderate distance of $d_*=10a$ (i.e. $\delta\simeq3.24$). It can be readily seen that there is an optimal swimmer-load size ratio in all cases.
Interestingly, Fig.~\ref{fig:Np8VelEff}a shows that while for moderate separation the dragging velocity decays with the increase in the load size, at close proximity it may actually be higher than the velocity of the unloaded swimmer. This is a direct consequence of Eq.~(\ref{eq:DragVelFixed}), which does not assume large separation: the fluid velocity in the center of the ``necklace" is larger than the swimming speed. This means, that in order to pull a load positioned at $d_*=0$ with the swimmer's speed, the applied force must act in the direction opposite to that of the velocity, so that $\lambda_l<0$ and $V_d>V_s$.

The numerical results confirm the qualitative dependencies arising from the far-field theory: there is a critical size-ratio $r_{cr}$ (weakly dependent on $\delta$) such, that for $r<r_{cr}$ the dragging efficiency is higher than the corresponding $\varepsilon_{d(\infty)}$ and for $r>r_{cr}$ the efficiency is lower than $\varepsilon_{d(\infty)}$; at moderate separations $r_{cr}\rightarrow \frac{1}{\sqrt{1+\varepsilon_s}}$ as expected from the far-field analysis. 
The discrepancy between the asymptotic and the numerical results is only observed at $r>1$, where the assumption $\frac{\delta}{r} \gg 1$ is no longer valid.

This work was supported by the Technion V.P.R. Fund. We thank J.~E. Avron for fruitful discussions.

\end{document}